\documentclass[aps,prx,twocolumn,superscriptaddress,longbibliography]{revtex4-2}
\usepackage{amssymb} 
\usepackage{graphicx,color} 
\usepackage{bbm} 
\usepackage{multirow}
\usepackage{amsmath}
\usepackage{mathrsfs} 
\usepackage{hyperref}
\usepackage[usenames,dvipsnames]{xcolor}
\usepackage{cancel}
\usepackage{color}
\definecolor{darkgreen}{rgb}{0,0.6,0}
\definecolor{darkblue}{rgb}{0,0,0.6}
\definecolor{darkred}{rgb}{0.6,0,0}
\definecolor{darkpurple}{rgb}{0.5,0,0.5}
\usepackage{soul}

\hypersetup{
bookmarksopen=true,
pdftitle="",
pdfauthor="", 
pdftoolbar=false, 
pdfstartview={FitH},		
pdfmenubar=true,			
pdfhighlight=/O,			
colorlinks=true,			
urlcolor=darkblue,
citecolor=darkblue,		
linkcolor=darkblue}

\usepackage{tikz}
 
\usepackage{amsmath}
\usepackage{amsfonts}
\usepackage{cases}
\usepackage{xspace}
\usepackage{psfrag}
\usepackage{color}
\usepackage{xcolor}

\newcommand{\ave}[1]{\mathchoice{\left\langle #1 \right\rangle}{\langle #1 \rangle}{\langle #1 \rangle}{\langle #1 \rangle}}

\newcommand{\plaind}{\mathrm{d}}
\newcommand{\dint}[1]{\mathchoice{\!\plaind#1\,}{\!\plaind#1\,}{\!\plaind#1\,}{\!\plaind#1\,}}

\usepackage{dsfont}

\newcommand{\canetset}[1]{{\mathchoice {\hbox{$\sf\textstyle #1\kern-0.4em #1$}}
{\hbox{$\sf\textstyle #1\kern-0.4em #1$}}
{\hbox{$\sf\scriptstyle #1\kern-0.3em #1$}}
{\hbox{$\sf\scriptscriptstyle #1\kern-0.2em #1$}}}}

\def\nbZ{{\mathchoice {\hbox{$\sf\textstyle Z\kern-0.4em Z$}}
{\hbox{$\sf\textstyle Z\kern-0.4em Z$}}
{\hbox{$\sf\scriptstyle Z\kern-0.3em Z$}}
{\hbox{$\sf\scriptscriptstyle Z\kern-0.2em Z$}}}}

\newcommand{\OC}{\mathcal{O}}

\newcommand{\ExpNB}[1]{\operatorname{exp}}
\renewcommand{\exp}[1]{\mathchoice{\mathrm{e}^{#1}}{\operatorname{exp}\left(#1\right)}{\operatorname{exp}\left(#1\right)}{\operatorname{exp}\left(#1\right)}}

\newcommand{\elabel}[1]{\label{eq:#1}}
\newcommand{\eref}[1]{(\ref{eq:#1})}
\newcommand{\Eref}[1]{Eq.~(\ref{eq:#1})}

\newcommand{\flabel}[1]{\label{fig:#1}}
\newcommand{\fref}[1]{Fig.~\ref{fig:#1}}

\newlength \standardfigwidth
\DeclareMathAlphabet{\matheub}{U}{eur}{m}{n}

\newcounter{exercise}
{\addtocounter{exercise}{1}\begin{center}\begin{minipage}{0.8\linewidth}\textbf{Exercise
\arabic{exercise}:}\begin{itshape}}
{\end{itshape}\end{minipage}\end{center}}

\makeatletter
\newcommand{\creat}[3][]{\@ifempty{#1}{#2^{\dagger}}{\left(#2^{\dagger}\right)^{#1}}\@ifempty{#3}{}{\!(#3)}}

\newcommand{\creatDoi}[3][]{\@ifempty{#1}{\tilde{#2}}{\left(\tilde{#2}\right)^{#1}}\@ifempty{#3}{}{(#3)}}

\newcommand{\annih}[3][]{#2\@ifempty{#1}{}{^{#1}}\@ifempty{#3}{}{(#3)}}

\makeatother

% ...

\newlength{\bibmarkkeyAleft}

\newlength{\bibmarkkeyBleft}

\newlength{\bibmarkkeyCleft}

\newlength{\bibmarkkeyDleft}

\usepackage{wasysym}

\makeatletter
\newcommand{\pushright}[1]{\ifmeasuring@#1\else\omit\hfill$\displaystyle#1$\fi\ignorespaces}
\newcommand{\pushleft}[1]{\ifmeasuring@#1\else\omit$\displaystyle#1$\hfill\fi\ignorespaces}
\makeatother

\newcommand{\diff}{\mathsf{D}_x}
\newcommand{\drift}{\mathsf{w}}

\newcommand{\pairPot}{W}

\usetikzlibrary{patterns}
\tikzset{
xxtsubstrate/.style={decorate, 
line width=1pt,
draw=cyan, 
decoration=snake, 
segment amplitude=0.75mm, 
line after snake=0.25mm,
line before snake=0.25mm
},
tsubstrate/.style={decorate, 
line width=1pt,
draw=cyan, 
decoration=snake, 
segment amplitude=0.5mm, 
segment length=5pt,
segment amplitude=0.2mm, 
line after snake=1mm,
line before snake=1mm
},
Bsubstrate/.style={decorate, 
line width=1pt,
draw=cyan, 
decoration=snake,
segment length=5pt,
segment aspect=0,
segment amplitude=0.5mm, 
line after snake=0mm,
line before snake=0mm
},
substrate/.style={decorate, 
line width=1pt,
draw=cyan, 
snake=snake, 
decoration=snake, 
segment length=5pt,
segment amplitude=0.5mm, 
line after snake=0.5mm,
line before snake=0.5mm
},
activity/.style={very thick,draw=red,postaction={decorate},
decoration={markings,mark=at position .5 with
{\arrow[draw=red]{>}}}},
cactivity/.style={very thick,draw=yellow,postaction={decorate},
decoration={markings,mark=at position .5 with
{\arrow[draw=yellow]{<}}}},
tactivity/.style={thick,draw=red,postaction={decorate},
decoration={markings,mark=at position .5 with
{\arrow[draw=red]{>}}}},
tEPSactivity/.style={thick,draw=red,postaction={decorate},
decoration={markings,mark=at position .55 with
{\arrow[draw=red]{>}}}},
tAactivity/.style={thick,draw=red},
Aactivity/.style={thick,draw=black},
Bactivity/.style={very thick,draw=black,dashed},
dotactivity/.style={very thick,draw=black,dotted},
Cactivity/.style={very thick,draw=black,decorate,decoration=snake},
Baractivity/.style={very thick,draw=black},
tSactivity/.style={thick,draw=red,postaction={decorate},
decoration={markings,mark=at position .7 with
{\arrow[draw=red]{>}}}},
Sactivity/.style={very thick,draw=red,postaction={decorate},
decoration={markings,mark=at position .7 with
{\arrow[draw=red]{>}}}}
}
\tikzset{
  terminal/.style = {
    rectangle,
    align = center,
    minimum size = 6mm,
    rounded corners = 3mm,
    very thick,
    draw = black!50,
    top color = white,
    bottom color = black!20,
  }
}

\tikzset{
tAsubstrate/.style={decorate, 
thick,
decoration=snake, 
segment amplitude=0.5mm, 
segment length=5pt,
segment amplitude=0.2mm, 
line after snake=1mm,
line before snake=1mm
},
DasHsubstrate/.style={draw=none,
postaction={decorate,decoration={markings,
        mark=at position 0.25 with {\draw[cyan] (0,-1.2mm) -- (0,1.2mm);}
        }}
},
tactivity/.style={thick,draw=red,postaction={decorate},
decoration={markings,mark=at position .5 with
{\arrow[draw=red]{>}}}},
tAactivity/.style={thick,draw=red},
tAactivityDasheD/.style={thick,draw=red,postaction={decorate,decoration={markings,
        mark=at position 0.75 with {\draw[red] (0,-1.2mm) -- (0,1.2mm);}
        }}},
DasheDtAactivity/.style={thick,draw=red,postaction={decorate,decoration={markings,
        mark=at position 0.25 with {\draw[red] (0,-1.2mm) -- (0,1.2mm);}
        }}},
tAactivityDA/.style={thick,draw=red,postaction={decorate,decoration={markings,
        mark=at position 0.6 with {\draw[red] (0,-1.2mm) -- (0,1.2mm);}
        }}},
AactivityDA/.style={thick,draw=black,postaction={decorate,decoration={markings,
        mark=at position 0.6 with {\draw[black,thin] (0,-1.2mm) -- (0,1.2mm);}
        }}},
DAtAactivity/.style={thick,postaction={decorate,decoration={markings,
        mark=at position 0.4 with {\draw (0,-1.2mm) -- (0,1.2mm);}
        }}},
potential/.style={thick,draw=black,dashed},
potentialDasheD/.style={thick, draw=black,densely dashed,postaction={decorate,decoration={markings,
        mark=at position 0.4 with {\draw[black,solid,thin] (0,-1.1mm) -- (0,1.1mm);}
        }}},
DasheDpotential/.style={thick,draw=black,densely dashed,postaction={decorate,decoration={markings,
        mark=at position 0.25 with {\draw[solid,thin] (0,-1.2mm) -- (0,1.2mm);}
        }}},
}

\tikzset{half paths/.style 2 args={%
  decoration={show path construction,
    lineto code={
      \draw [#1,out=150,in=30] (\tikzinputsegmentfirst) to[bend right]
         ($(\tikzinputsegmentfirst)!0.5!(\tikzinputsegmentlast)$);
      \draw [#2,out=150,in=30] ($(\tikzinputsegmentfirst)!0.5!(\tikzinputsegmentlast)$)
        to[bend right] (\tikzinputsegmentlast);
    }
  }, decorate
}}

\newcommand{\LoopDiagramSmall}[9]{
    \tikz[baseline=-2.5pt,scale=0.65]{
    \begin{scope}[xshift=-1.4cm]
    \draw[#1] (-0.4,0.4) -- (0,0.3);
    \draw[draw=none,in=30,out=150]  (-0.4,0.4) -- node[pos=0.4,above=0.1cm] {} 
    node[pos=0.4,sloped] {\tikz{\draw[#2] (0,-1.2mm) -- (0,1.2mm);}} (0,0.3);
    \draw[DasheDpotential] (0,0.3) -- (0,-0.3);
    \draw[#6] (-0.4,-0.4) -- (0,-0.3);
    \end{scope}
    \draw[pattern=north west lines] (0,0) circle (0.7cm);
    \draw[fill=white] (0,0) node[circle,fill=white,inner sep=2pt,opacity=0.9] {$n$}; 
    \draw[#3,out=150,in=180] (150:0.7cm) to[bend right] (-1,0.501);
    \draw[#1,out=180,in=30] (-1,0.501) to[bend right] (-1.4cm,0.3);
    \draw[draw=none,in=180,out=150]  (150:0.7cm) to[bend right] node[pos=0.4,above=0.1cm] {$#5$} 
    node[pos=0.4,sloped] {\tikz{\draw[#4] (0,-1.2mm) -- (0,1.2mm);}} (-1,0.501);
    \draw[#7,in=-180,out=210] (210:0.7cm) to[bend left] (-1cm,-0.501);
    \draw[#6,out=180,in=-30] (-1cm,-0.501) to[bend left] (-1.4cm,-0.3);
    \draw[draw=none,in=-30,out=210]  (210:0.7cm) to[bend left] node[pos=0.4,below=0.1cm] {$#9$} 
    node[pos=0.4,sloped] {\tikz{\draw[#8] (0,-1.2mm) -- (0,1.2mm);}} (-1,-0.501);
    }
}

\tikzset{
    position label/.style={
       below = 3pt,
       text height = 1.5ex,
       text depth = 1ex
    },
   brace/.style={
     decoration={brace, mirror},
     decorate
   }
}

\usetikzlibrary{patterns}
\usetikzlibrary{decorations.pathmorphing}
\usetikzlibrary{decorations.markings}

  \usetikzlibrary{snakes}

\usepackage{CJK}

\newcommand{\tildephi}{\widetilde\phi}
\newcommand{\tildepsi}{\widetilde{\psi}}

\allowdisplaybreaks

\newcommand{\titleText}{Effective attraction by repulsion}
\title{\titleText}

\CJKnospace
\newcommand{\zilname}{\begin{CJK*}{UTF8}{gbsn}张子洛\end{CJK*}}

\begin{document}

\title{\titleText}
\author{Rosalba Garcia-Millan}
\affiliation{Department of Mathematics, King's College London, Strand, London WC2R 2LS, United Kingdom}
\affiliation{DAMTP, Centre for Mathematical Sciences, University of Cambridge, Cambridge CB3 0WA, United Kingdom}
\affiliation{St John's College, University of Cambridge, Cambridge CB2 1TP, United Kingdom}
\author{Luca Cocconi}
\affiliation{Department of Mathematics and Centre of Complexity Science, Imperial College London, London SW7 2AZ, United Kingdom}
\author{Ziluo Zhang (\zilname)}
\affiliation{Department of Mathematics and Centre of Complexity Science, Imperial College London, London SW7 2AZ, United Kingdom}
\author{Marius Bothe}
\affiliation{Department of Mathematics and Centre of Complexity Science, Imperial College London, London SW7 2AZ, United Kingdom}
\author{Letian Chen}
\affiliation{Department of Mathematics and Centre of Complexity Science, Imperial College London, London SW7 2AZ, United Kingdom}
\author{Zigan Zhen}
\affiliation{Department of Mathematics and Centre of Complexity Science, Imperial College London, London SW7 2AZ, United Kingdom}
\author{Gunnar Pruessner}
\email{g.pruessner@imperial.ac.uk}
\affiliation{Department of Mathematics and Centre of Complexity Science, Imperial College London, London SW7 2AZ, United Kingdom}
\date{\today}

\begin{abstract}
Repulsive self-propelled particles tend to cluster, 
leading to Motility-Induced Phase Separation (MIPS). 
By analogy with equilibrium phase separation, the onset of MIPS has been associated 
with a transition to effective attraction between particles. 
Using an exact microscopic theory, we  quantify the emergence of effective attraction 
in a minimal model: two soft run-and-tumble particles
in a periodic domain.
We show that, as repulsion increases, the leading-order behaviour is that of effective repulsion, while effective attraction emerges as a higher-order contribution to the renormalisation of the 
pair potential.
\end{abstract}

\maketitle

\newcommand{\xe}{x_{\text{e}}}
\newcommand{\xo}{x_{\text{o}}}
\newcommand{\xJ}{x_{\text{J}}}
\newcommand{\xA}{x_{\text{A}}}
\newcommand{\xRJ}{x_{\text{RJ}}}
\newcommand{\Pe}{\mathsf{Pe}}
\newcommand{\xibar}{\bar{\xi}}
\newcommand{\nubar}{\bar{\nu}}
\newcommand{\gammabar}{\bar{\gamma}}
\newcommand{\dimensionlessParams}[1]{\textcolor{darkgreen}{[#1]}}

\newcommand{\Prob}[1]{P_{\hspace{-2pt}{\scalebox{.8}{$\scriptscriptstyle #1$}}}}
\newcommand{\Ppp}{\Prob{++}}
\newcommand{\Ppm}{\Prob{+-}}
\newcommand{\Pmp}{\Prob{-+}}
\newcommand{\Pmm}{\Prob{--}}
\newcommand{\Pss}{\Prob{\sigma_1 \sigma_2}}

\newcommand{\Weff}{\pairPot_{\hspace{-1pt}\text{eff}}}

Self-propelled agents defy thermodynamic equilibrium by locally converting
chemical energy into persistent motion,
giving rise to a vast range of intriguing phenomena, such as phase separation in the absence of attraction \cite{FilyMarchetti:2012,
CatesTailleur:2015,
DigregorioETAL:2018,
CatesNardini:2025,
VanDerLindenETAL:2019,
ZhangETAL:2021},
flocking 
\cite{CavagnaGiardina:2014,BarberisPeruani:2016,PaoluzziLevisPagonabarraga:2024,KumarETAL:2014},
and spontaneous particle rectification in ratcheted landscapes
\cite{ReinETAL:2023,ZhenPruessner:2022,DiLeonardoETAL:2025,ReichhardtReichhardt:2017,ReinETAL:2023}. 
Understanding the mechanisms underlying these phenomena is of fundamental importance
to uncover the principles 
of nonequilibrium physics across scales.

Free active particles experience enhanced long-time diffusion, which may be described by a higher effective temperature 
\cite{BoriskovskyLindnerRoichman:2024,
BoriskovskyETAL:2025,
PalacciETAL:2010,
LevisBerthier:2015,
Szamel:2014}. 
In apparent contrast, collections of active particles that interact repulsively at short distances
tend to accumulate, forming
dense clusters in otherwise dilute regions where particles
move freely and
barely interact. This phenomenon is known as Motility-Induced Phase Separation (MIPS)
and
originates from the feedback between
particle slowdown and accumulation 
\cite{CatesNardini:2025,
CatesTailleur:2015}.
MIPS has been a primary focus in nonequilibrium statistical mechanics for over a decade now, 
as it represents a genuinely nonequilibrium phase transition, one that is clearly forbidden by the laws of equilibrium thermodynamics. In particular, given the equilibrium tenet of ``no phase separation (of a one component fluid) without pair attraction'', it is no surprise that the onset of MIPS has 
been attributed to
the \emph{emergence of effective attraction} between particles
\cite{SolonCatesTailleur:2015,
WittkowskiETAL:2014}. 
So far, however, this equilibrium analogy has lacked an analytical grounding.  

\begin{figure}
\includegraphics[width=0.49\textwidth]{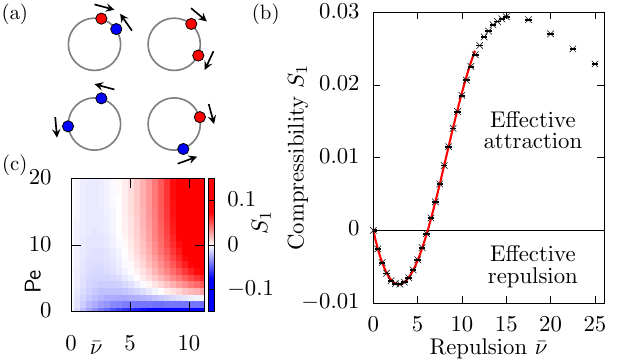}
\caption{\flabel{cosmode} 
(a) Two soft RTPs in a periodic domain (arrows indicate orientation).
(b)
Compressibility factor $S_1=\ave{\cos(k_1x)}$, as a
function of repulsion $\nubar$,
showing the crossover from effective repulsion ($S_1<0$) to effective attraction ($S_1>0$).
Analytical results (line) and numerical results (symbols) agree.
Parameters: $\diff=2$, $L=20$, 
{$\xibar=0.01$, $\Pe=2$, $\gammabar=0.02$}.
(c) 
Compressibility 
as a function of $\nubar$ and $\Pe$. 
Blue indicates effective repulsion, red indicates effective attraction.
Parameters: $\diff=1$, $L=20$, $\xibar=0.02$, $\gammabar=0.02$. 
}
\end{figure}

Instead, existing theoretical descriptions of MIPS largely
follow a top-down approach, where bare interactions, or particle-particle interactions,
are approximated by phenomenological particle-density interactions \cite{CatesNardini:2025,
CatesTailleur:2015,WittkowskiETAL:2014}. These may in turn be coarse grained into effective terms in a continuous-density theory, typically a nonequilibrium generalisation of Model B 
\cite{CatesNardini:2025,
CatesTailleur:2015,
TjhungNardiniCates:2018,
YadavMishraPuri:2025}.
Such coarse-grained descriptions successfully predict the instability of the homogeneous phase
towards a spinodal decomposition into regions of two distinct
densities separated by sharp boundaries \cite{CatesTailleur:2015}.
However, as a compromise stemming from their multiscale nature, 
the predictive power of these descriptions comes at the expense of no clear connection with microscopic mechanisms \cite{CatesNardini:2025}, particularly the origin of effective pair interactions.
Soto et al.~\cite{SotoPintoBrito:2024} offer a perspective 
via a Boltzmann-Enskog-like treatment of pair collisions in the high persistence regime.
Starting from active lattice models,
other studies derive
hydrodynamic equations for coarse-grained variables
evolving on length scales large compared to the interaction range of individual particles
\cite{MasonETAL:2023,
LiETAL:2023,
Kourbane-HousseneETAL:2018,
MukherjeeETAL:2025,
BaskaranMarchetti:2008,
MarchettiETAL:2013,
MartinETAL:2021,
FodorETAL:2016}.

To elucidate the role of persistence, repulsion and reorientation
in the emergence of particle accumulation and to clarify the extent to which this mechanism may be subsumed under an effective attraction, it is essential to 
retain microscopic details in a theoretical description.
In this spirit, we consider
two soft run-and-tumble particles (RTPs)
in a continuous, one-dimensional, periodic space, \fref{cosmode}(a).
Following a bottom-up approach
\cite{BialkeLoewenSpeck:2013,
SotoPintoBrito:2024}, we derive an exact microscopic theory 
and characterise the nonequilibrium stationary state analytically.
The main observable of interest in this work is the two-point correlation function, whose logarithm
defines the effective potential.
We use the lowest non-zero Fourier mode of the correlation function, 
essentially the compressibility factor, to quantify effective interactions 
Figs.~\ref{fig:cosmode}(b) and (c)
\footnote{Strictly, the observable $S_1=\ave{\cos(k_1x)}$ is the shifted, rescaled
compressibility $\chi_T=-\frac{1}{V}\left(\frac{\partial V}{\partial P}\right)_T$
of a volume $V$ at pressure $P$ and constant temperature $T$,
according to
$\chi_T L\rho_0^2 k_B T /2= 1 + S_1$ \cite{HansenMcDonald:2006}.}.
While both hard and soft active particles can exhibit MIPS \cite{SanoriaETAL:2021,
SanoriaETAL:2024}, seminal studies have predominantly dealt with the rather more tractable case of RTPs interacting via hard-core exclusion, specifically
on ring-lattices
\cite{SlowmanETAL:2016,
SlowmanETAL:2017,
MallminETAL:2019,
GuillinHahnMichel:2025}
or in the continuum
\cite{DasDharKundu:2020,
HahnGuillinMichel:2025,
MetsonEvansBlythe:2023,
preprintHahnGuillinMichel:2025}.
As we show in this work, new physics arises from soft repulsion that is
hidden in (singular) hard-core exclusion \cite{WillemsETAL:2025}:
we find that
the active system crosses over from a repulsive effective potential to
an attractive one by increasing
bare particle-particle repulsion, \fref{cosmode}(b).

To compute correlation functions, we follow a
path-integral approach in the framework of
Doi-Peliti field theory (DPFT) 
\cite{Doi:1976,Peliti:1985},
where the action derives from a second quantisation of a 
Fokker-Planck equation \cite{PruessnerGarcia-Millan:2025}.
Crucially, the formalism retains particle entity and thus captures the
conservative-multiplicative noise inherent in particle systems exactly
\cite{BotheETAL:2023}.
The key technical challenge concerns the renormalisation of bare interactions
into effective interaction vertices, which we construct as a perturbation expansion in the interaction coupling.
Our calculation systematically accounts for contributing terms
from the aligned and anti-aligned configurations, 
showing how these ``feed off'' each other as a consequence of particles tumbling. 
Diagrammatically, our work is reminiscent of Feynman's
scattering theory \cite{Feynman:1949,PeskinSchroeder:95},
highlighting the role of field theory as the go-to language
to describe effective interactions between point particles \cite{ZhangGarcia-Millan:2023}.

\emph{Model} --- We consider two indistinguishable RTPs
on a periodically closed interval of length $L$ at positions $x_i\in[-L/2,L/2)$, $i\in\{1,2\}$,
governed by the overdamped Langevin equations
\begin{equation}
\dot{x}_i =	 \drift\sigma_i(t) - \pairPot'(x_i-x_j) + \sqrt{2\diff}\eta_i(t) \ .
\elabel{OLE}
\end{equation}
The self-propulsion has constant magnitude $\drift$
and fluctuating orientation $\sigma_i(t)\in\{-1,1\}$ 
that switches between values with Poissonian rate
$\gamma$, \fref{cosmode}(a).
The thermal noise $\eta_i(t)$ is Gaussian with mean $\ave{\eta_i(t)}=0$ and variance
$\ave{\eta_i(t)\eta_j(t')} = \delta_{i,j}\delta(t-t')$.
We use a Yukawa potential $\propto \exp{-|x|/\xi}$ to model soft interaction
forces $\pairPot'$ at distance $x=x_1-x_2$,
with interaction length $\xi$ \cite{TADPvanWijland:2023}.
On a periodic space, 
the Yukawa potential is
\begin{equation}
    \pairPot(x) = \frac{\nu \cosh(|x|/\xi-L/(2\xi))}{2\xi \sinh(L/(2\xi))} \ ,
    \elabel{pair_potentialRing}
\end{equation}
where $\nu$ is the interaction coupling,
normalised such that $\int_{-L/2}^{L/2}\dint{x} \pairPot(x) = \nu$.
Where convenient, we use the
dimensionless parametrisation
$\xibar = {\xi}/{L}$,
$\nubar = {\nu}/{(\diff\xi)}$,
$\Pe= {\drift^2}/{(\diff\gamma)}$, and
$\gammabar  = {\gamma \xi^2}/{\diff}$.

Particles colliding head-on tend to jam at
a finite distance from each other. This distance, where particle velocities are zero on average,
is what we call the \emph{jamming distance} $\xJ$.
Aggregation is often explained at a coarse-grained level as a self-reinforcing mechanism whereby
particles slow down where they accumulate
and accumulate where they slow down
\cite{CatesNardini:2025,
CatesTailleur:2015,
SolonCatesTailleur:2015}. 
Setting aside 
that this suggests particle accumulation
in monodisperse equilibrium systems with a suitable initial configuration (an unphysical scenario), 
there is no microscopic theory that demonstrates the 
link between particle slowdown and particle accumulation.
In fact, as we show below, the distances where particles jam and where they accumulate are different.

Rewriting the dynamics \eref{OLE}  
in the co-moving frame of either particle gives
the equation of motion
of the inter-particle distance $x$
in the aligned 
and anti-aligned 
configurations,
\begin{equation}
\elabel{OLE_naive}
\dot{x} = \left\{
\begin{aligned}
&- 2\pairPot'(x) + \sqrt{2\diff}(\eta_1 - \eta_2)   && \text{ if }\sigma_1=\sigma_2 \ ,\\
&2\drift\sigma - 2\pairPot'(x) + \sqrt{2\diff}(\eta_1 - \eta_2)  && \text{ if }\sigma_1=-\sigma_2\equiv\sigma \ ,
\end{aligned}
\right.
\end{equation}
which can be interpreted as a Brownian motion
switching between total potentials $2\pairPot(x)$ and $2(-\drift \sigma x + \pairPot(x))$
with rate $\gamma$ \cite{Risken:1996}.
When short-range repulsion exceeds the self-propulsion force, 
anti-aligned particles that encounter each other can transiently establish a bound state at distance $\xJ$,
where the tilted potential has a minimum $\drift\sigma = \pairPot'(\xJ)$. 
Assuming $\xi\ll L$, the jamming distance
$\xJ$ is approximated by
\begin{equation}
\elabel{jammingDist_naive}
\frac{\xJ}{\xi} \simeq \ln\left(\frac{\nu}{2\drift\xi^2}\right) 
\end{equation}
provided  $\nu\geq2\drift\xi^2 $.
The distance $\xJ$ does not incorporate the tumbling and therefore fails to capture the emergent
bound state if the relaxational timescale of \Eref{OLE_naive} exceeds the time between tumbles. 

\emph{Microscopic field theory} --- We derive a DPFT
that captures soft interactions 
as well as tumbling in the dynamics of \Eref{OLE} systematically.
Here, we present the key steps of our derivation and associated Feynman diagrams, 
and leave the technical details to \cite{companion}.
We represent
right- and left-moving particles with field pairs $\phi,\tildephi$ and $\psi,\tildepsi$ respectively
\cite{PruessnerGarcia-Millan:2025,
Garcia-MillanPruessner:2021,
RobertsPruessner:2022,
ZhangPruessner:2022}.
Each species has associated with it an annihilation field $\phi$, $\psi$, 
and a Doi-shifted creation field
$\tildephi$, $\tildepsi$. We obtain
four bare propagators governing free motion: same-species propagators
accounting for propulsion, diffusion
and an even number of reorientations,
\begin{align}
\elabel{bare_props1}
\tikz[baseline=-2.5pt,scale=0.8]{
\draw[Aactivity] (0,0) -- (1,0) node [at start, above] {$\phi$} node [at end, above] {$\tildephi$};
} 
&& \text{and} &&
\tikz[baseline=-2.5pt,scale=0.8]{
\draw[tAsubstrate] (0,0) -- (1,0) node [at start, above] {$\psi$} node [at end, above] {$\tildepsi$};
} \ ;
\end{align}
and transmutation propagators
involving an odd number of reorientations,
\begin{align}
\elabel{bare_props2}
\tikz[baseline=-2.5pt,scale=0.8]{
\draw[tAsubstrate] (0,0) -- (0.5,0) node [at start, above] {$\psi$};
\draw[Aactivity] (0.5,0) -- (1,0) node [at end, above] {$\tildephi$};
} 
&& \text{and} &&
\tikz[baseline=-2.5pt,scale=0.8]{
\draw[Aactivity] (0,0) -- (0.5,0) node [at start, above] {$\phi$};
\draw[tAsubstrate] (0.5,0) -- (1,0) node [at end, above] {$\tildepsi$};
}  \ .
\end{align}
Particle interaction mediated by a pair potential  is captured
by the four-point vertices
\begin{align}
\elabel{vertices_main}
 \tikz[baseline=-2.5pt,scale=0.8]{
    \draw[Aactivity] (-0.4,0.4) -- (0,0.3) -- (0.4,0.4);
    \draw[draw=none] (-0.4,0.4) -- 
      node[pos=0.6,sloped] {\tikz{\draw (0,-1.2mm) -- (0,1.2mm);}} (0,0.3) ;
    \draw[DasheDpotential] (0,0.3) -- (0,-0.3);
    \draw[Aactivity] (-0.4,-0.4) -- (0,-0.3) -- (0.4,-0.4);
    }
 \  , &&
     \tikz[baseline=-2.5pt,scale=0.8]{
    \draw[tAsubstrate] (-0.4,0.4) -- (0,0.3) -- (0.4,0.4);
    \draw[draw=none] (-0.4,0.4) -- 
      node[pos=0.6,sloped] {\tikz{\draw (0,-1.2mm) -- (0,1.2mm);}} (0,0.3) ;
    \draw[DasheDpotential] (0,0.3) -- (0,-0.3);
    \draw[tAsubstrate] (-0.4,-0.4) -- (0,-0.3) -- (0.4,-0.4);
    } 
 \  , &&
    \tikz[baseline=-2.5pt,scale=0.8]{
    \draw[Aactivity] (-0.4,0.4) -- (0,0.3) -- (0.4,0.4);
    \draw[draw=none] (-0.4,0.4) -- 
      node[pos=0.6,sloped] {\tikz{\draw (0,-1.2mm) -- (0,1.2mm);}} (0,0.3) ;
    \draw[DasheDpotential] (0,0.3) -- (0,-0.3);
    \draw[tAsubstrate] (-0.4,-0.4) -- (0,-0.3) -- (0.4,-0.4);
    }
 && \text{and} &&
    \tikz[baseline=-2.5pt,scale=0.8]{
    \draw[tAsubstrate] (-0.4,0.4) -- (0,0.3) -- (0.4,0.4);
    \draw[draw=none] (-0.4,0.4) -- 
      node[pos=0.6,sloped] {\tikz{\draw (0,-1.2mm) -- (0,1.2mm);}} (0,0.3) ;
    \draw[DasheDpotential] (0,0.3) -- (0,-0.3);
    \draw[Aactivity] (-0.4,-0.4) -- (0,-0.3) -- (0.4,-0.4);
    } \ ,
\end{align}
each proportional to the Yukawa interaction coupling $\nubar$
\cite{PruessnerGarcia-Millan:2025,ZhangGarcia-Millan:2023,companion}. 
Interpretation of the interaction vertices 
in \eref{vertices_main}
is as follows:
the third diagram,
for instance, implements the effect of the interaction on the motion of a right-moving particle
(\tikz[baseline=-2.5pt,scale=0.5]{
    \draw[Aactivity]  (0,0) -- (1,0);
    \draw[draw=none] (1,0) --  node [pos=0.3]{\tikz{\draw (0,-1.2mm) -- (0,1.2mm);}} (0,0) ;
})
due to the presence of a left-moving particle
(\tikz[baseline=-2.5pt,scale=0.5]{
\draw[tAsubstrate] (0,0) -- (1,0);
})
quantified by the force
(\tikz[baseline=-2.5pt,scale=0.5]{
    \draw[potentialDasheD] (0,0.3) -- (0,-0.3);
    }).

\begin{figure}
\includegraphics[width=0.48\textwidth]{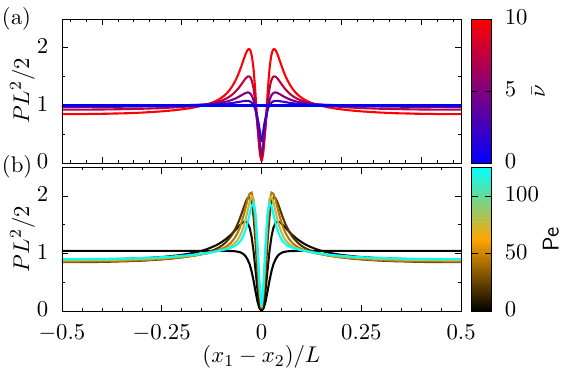}
\caption{\flabel{stat_corr_func}
Joint particle density 
$P(x_1,x_2)$ as a function of relative particle distance
varying (a) repulsion $\nubar$ and (b) activity $\Pe$.
Parameters: $\diff=0.5$, $L=20$, $\gammabar=0.008$, $\xibar=0.01$, (a)  $\Pe=20$, and (b) $\nubar=10$.
}
\end{figure}

The \emph{effective} interaction vertices fully
capture emergent effective interaction and
are renormalised by bare interaction, activity and tumbling.
The effective interaction vertices are constructed as
the sum
of all loop corrections in the perturbation expansion about the Yukawa coupling $\nubar$ that need to be added to 
the bare interaction
vertices in \eref{vertices_main}.
Under symmetries in the stationary state, all effective vertices reduce to two,
    \tikz[baseline=-2.5pt,scale=0.5]{
    \draw[pattern=north west lines] (0,0) circle (0.7cm);
    \draw[DAtAactivity] (150:0.7cm) -- (150:1.25cm);
    \draw[draw=none] (150:0.7cm) -- (150:1.05cm) ;
    \draw[Aactivity] (210:0.7cm) -- (210:1.25cm); 
}    
and
    \tikz[baseline=-2.5pt,scale=0.5]{
    \draw[pattern=north west lines] (0,0) circle (0.7cm);
    \draw[DAtAactivity] (150:0.7cm) -- (150:1.25cm);
    \draw[draw=none] (150:0.7cm) -- (150:1.05cm) ;
    \draw[tAsubstrate] (210:0.7cm) -- (210:1.25cm); 
},
 corresponding to the aligned and anti-aligned configurations
respectively.
Calculating the effective vertices analytically
is the fundamental challenge of this problem.
Here, we use an iterative scheme whereby the $(n+1)$-th order in $\nubar$ is
calculated systematically by adding a loop correction
generated with the vertices in \eref{vertices_main}
to the $n$-th order  \cite{companion}, 
for instance,
\begin{align}
    \tikz[baseline=-2.5pt,scale=0.75]{
    \draw[pattern=north west lines] (0,0) circle (0.7cm);
    \draw[DAtAactivity] (150:0.7cm) -- (150:1.15cm);
    \draw[draw=none] (150:0.7cm) -- (150:1.05cm) ;
    \draw[tAsubstrate] (210:0.7cm) -- (210:1.15cm); 
    \draw[fill=white] (0,0) node[circle,fill=white,inner sep=1pt,opacity=0.9] {\small{$n+1$}};
    }
 & = 
\LoopDiagramSmall{Aactivity}{}{Aactivity}{}{}{tAsubstrate}{Aactivity}{draw=none}{}
\ +
\LoopDiagramSmall{Aactivity}{}{Aactivity}{draw=none}{}{tAsubstrate}{Aactivity}{}{}
\ +
\LoopDiagramSmall{Aactivity}{}{tAsubstrate}{}{}{tAsubstrate}{Aactivity}{draw=none}{}
 \nonumber\\
&+
\LoopDiagramSmall{Aactivity}{}{tAsubstrate}{draw=none}{}{tAsubstrate}{Aactivity}{}{}
\ +
\LoopDiagramSmall{Aactivity}{}{Aactivity}{}{}{tAsubstrate}{tAsubstrate}{draw=none}{}
\ +
\LoopDiagramSmall{Aactivity}{}{Aactivity}{draw=none}{}{tAsubstrate}{tAsubstrate}{}{}
 \nonumber\\
&+
\LoopDiagramSmall{Aactivity}{}{tAsubstrate}{}{}{tAsubstrate}{tAsubstrate}{draw=none}{}
\ +
\LoopDiagramSmall{Aactivity}{}{tAsubstrate}{draw=none}{}{tAsubstrate}{tAsubstrate}{}{}
\elabel{tHeteroPot_recurrence}
\ .
\end{align}
Calculating observables, such as correlation functions,
is then a matter of attaching the suitable outgoing legs to the
effective interaction vertices \cite{companion}.

\begin{figure*}
\includegraphics[width=\textwidth]{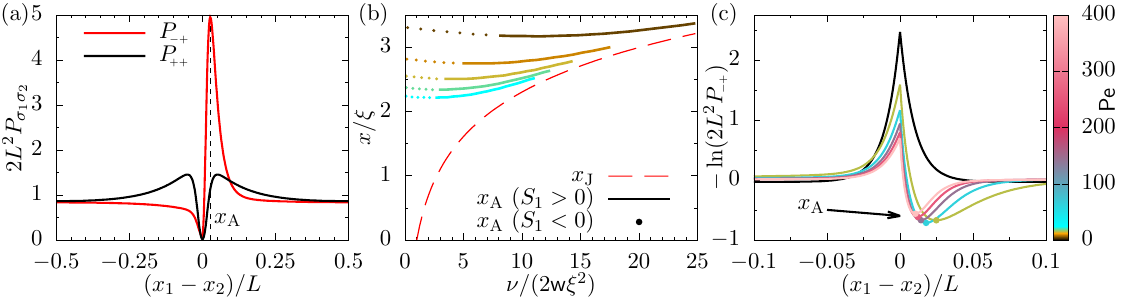}
\caption{\flabel{stat_corr_func_partial} 
(a) Stationary joint two-point correlation functions $P_{++}$
and $P_{-+}$.
Parameters: $\diff=0.5$, $L=20$, $\xibar=0.01$, 
$\nubar=10$, $\Pe=20$, $\gammabar=0.008$.
\flabel{stat_corr_func_compare} 
(b) 
Accumulation distance $\xA$ (dots and lines),
the maximum of $P_{-+}$ as shown in (a),
and jamming distance $\xJ$ (dashed), \Eref{jammingDist_naive}, as a function of 
$\nu$.
Dots indicate effective repulsion ($S_1<0$) and
solid lines indicate effective attraction ($S_1>0$).
Parameters: $\diff=1$, $L=20$, $\xibar=0.01$, $\gammabar=0.01$.
(c)
Effective potential $\Weff=-\ln(2L^2P_{-+})$
between right- and left-moving particles.
Parameters: $\diff=1$, $L=20$, $\nubar=5$, $\xibar=0.01$, $\gammabar=0.01$.
\flabel{jammingDist}
}
\end{figure*}

\emph{Emergence of effective attraction} --- The stationary
two-point correlation function $P(x_1,x_2)$,
or joint particle number density 
\cite{HansenMcDonald:2006,FarageETAL:2015}, is shown in
\fref{stat_corr_func}(a) for varying $\nubar$.
We find that,
as the repulsion is increased, the profile $P$ smoothly departs from the uniform distribution
$P = {2!}/{L^2}$ at $\nubar=0$
acquiring two features: first, a minimum at $x_1=x_2$, signature of short-range repulsion,
and second, the growth of two symmetric local maxima indicating 
two emergent bound states.
\fref{stat_corr_func}(b) shows $P$
for increasing activity $\Pe$,
as it deviates from the purely repulsive equilibrium Boltzmann distribution $P \propto \exp{- \pairPot(x)/\diff}$
at $\Pe=0$
by exhibiting the two emergent bound states.

In a perturbation theory, even in equilibrium, the effect of short-ranged attraction is 
exponentially enhanced as particles spend more time close together interacting more strongly, whereas the effect of repulsion is exponentially suppressed as they are separated thereby interacting less. This is reflected in a perturbation expansion of any observable in coupling strength $\nubar$, as, order by order, the sign of $(-\nubar)^n$ is always positive in the case of attraction, $\nubar<0$, while it alternates for repulsion. Attraction is therefore self-reinforcing, while repulsion is self-limiting. In the present case, this mechanism favours effective attraction over bare repulsion. A second mechanism at play is the increase of effective diffusivity $\mathsf{D}_{\text{eff}}=\diff(1+\Pe/2)$  due to activity \cite{ZhangPruessner:2022}.
In the observable considered below, the two combine to reduce the amplitude of odd powers of $\nubar$, while maintaining the amplitude of even powers. 
However, drawing intuition from the passive case is somewhat misguided, because the phenomenon of effective attraction reported here is inherently dynamical, as the particles permanently change relative orientation.

We use the compressibility factor $S_1=\ave{\cos(k_1 x)}$ 
with wavenumber $k_1=2\pi/L$ to
quantify effective interactions
\cite{FilyMarchetti:2012,HansenMcDonald:2006}.
The compressibility factor
measures the weight of $P$ over 
short distances $0\leq |x| \leq L/4$ in relation to its weight over long distances 
$L/4 \leq |x| \leq L/2$.
We consider $S_1>0$ as the fingerprint of effective attraction
and $S_1<0$ of effective repulsion.
For the non-interacting case, 
the compressibility factor trivially is $S_1=0$.
In the passive limit, the compressibility is 
$S_1= \xibar\sum_{n\geq1}{(-\nubar)^n }/({2^{n-2}(n-1)!(k_1^2\xi^2+n^2)})
$,
negative for any repulsive $\nubar>0$.

In the interacting, active case, our analytical results show that 
\begin{align}
S_1 = & -  \frac{\nubar \xibar \left(2 \left(k_1^2\xi^2+\gammabar \right) \left(k_1^2\xi^2+ 2\gammabar\right) +\Pe \gammabar k_1^2\xi^2 \right)}{\left( k_1^2\xi^2+1 \right) \left(k_1^2\xi^2+\gammabar \right) \left(k_1^2\xi^2+\gammabar(2+\Pe) \right)}
 +\OC\left(\nubar^2\right) 
 \ .
\end{align}
The sign indicates that the leading-order behaviour in $\nubar$
 is that of effective repulsion,
irrespective of the activity $\Pe$ and any other system parameter, \fref{cosmode}(b)
 \cite{companion}.
Thus, for small couplings $\nubar$, repulsive bare interactions
result in effective repulsion between them, though this is reduced by the activity.
We find that as $\nubar$ is increased further, higher-order terms become relevant,
eventually rendering $S_1$ positive. 
Therefore, \emph{soft particle-particle repulsion needs to be sufficiently strong for the
particles to display effective attraction}.
Our iterative scheme generates higher-order corrections analytically but these are
generally cumbersome and are thus relegated to the companion paper \cite{companion}.

The crossover from effective repulsion to effective attraction in \fref{cosmode}(b) takes place around 
$\nubar\simeq6.25$, 
which is an order of magnitude larger than the smallest $\nubar$ that meets 
the existence condition $\nubar=2\sqrt{\Pe\gammabar}=0.4$ for $\xJ$. 
This exemplifies how, although the simple force-balance argument presented above
provides an explanation for the jamming
mechanism,
its prediction does not extend to effective interactions, because it neglects orientation
dynamics.
The sign of the compressibility factor $S_1$ 
yields the phase diagram in \fref{cosmode}(c) showing that low $\nubar$ and $\Pe$ result in 
effective repulsion (blue), whereas effective attraction (red) emerges at high enough $\nubar$ and $\Pe$.

The role of active fluctuations, in this case
induced by tumbling,
in the emergence
of effective interactions becomes apparent in the pair correlation functions
that account for the internal state of either particle.
These are the joint probability densities $\Pss(x_1,x_2)$ that 
particles are at positions $x_1$, $x_2$
and have internal states $\sigma_1$, $\sigma_2$.
By marginalisation over the internal states, we recover the 
joint particle number density
above, 
\begin{align}
P(x_1,x_2) = & \Ppp(x_1,x_2) + \Ppm(x_1,x_2) \nonumber \\ & + \Pmm(x_1,x_2) + \Pmp(x_1,x_2) \ .
\end{align}
Under symmetries in the stationary state, the right-hand side can be expressed in terms of $\Ppp$ and $\Pmp$, 
which are shown in \fref{stat_corr_func_partial}(a).

Tumbling resets particles in a new configuration of internal states
at a transient, unstable distance, 
so that distributions feed into each other, say
$\Ppp$ into $\Pmp$ and vice versa.
The relaxational timescale at which this happens is central to the effect this has. 
When particles change orientation, relaxing from the anti-aligned state of a head-on collision 
at distance $\xJ$ to the distribution of aligned particles takes about $\propto L^2/\diff$, as $L$ is
the available space and the distance $\xJ$ is much smaller than that. 
When they change back to anti-aligned, it 
takes about $\propto L/\drift$ to return to being side by side at distance $\xJ\ll L$. 
The particles stay at a distance provided the potential barrier $\Delta E = W(0)-W(\xJ)\simeq \nu/(2\xi) - \drift\xi$
is not overcome before they tumble again, $1/\gamma> \hspace{2pt}\sim\hspace{-4pt}\exp{\Delta E/\diff}$.
A switching rate $ \gamma$ such that $ L^2/\diff \gtrsim 1/\gamma \gtrsim L/\drift$ will allow particles to spend much time close to each other in the anti-aligned state and 
in the aligned state turn around before they diffuse apart. 
As a result, even aligned particles, whose relative dynamics is purely diffusive and whose interaction is 
purely repulsive,
\Eref{OLE_naive}, stay together over relatively long times as if attracted to each other.
If particles do not have enough time to relax between consecutive reorientations, this 
mechanism of resetting at close distances 
is reinforced over time. 
Similarly, particles in the anti-aligned configuration that separate from each other
do not separate for long before one of them tumbles.
Even if none of them tumbles, they are bound to meet again due to the system's periodicity.
With tumbling, particles do not jam for so long that they overcome the potential barrier, 
but when they start separating, they also do not separate for so long that they attain the equilibrium distribution.
This  back-and-forth dynamics reinforces effective attraction
between particles in either internal-state configuration.
Analytically, this mixing is captured in our iterative scheme \cite{companion},
which determines how the densities
$\Ppp$ and $\Pmp$ feed into each other order by order in the
perturbation expansion in $\nubar$.
We find that the part of $\Pmp$ odd in $x$
arises at $\Pe>0$ for orders $n\geq2$ in $\nubar$, which is denoted as $R_n$ in 
\cite{companion} and is the signature of activity.
This term $R_n$ suppresses the amplitude of odd orders in the perturbation expansion of the
compressibility factor $S_1$ in $\nubar$, leading to an effective attraction as indicated by $S_1>0$.

The correlation function $\Ppp$, shown in \fref{stat_corr_func_partial}(a)
has
two apparent ``bound" states between aligned particles due to
the feeding from the anti-aligned configuration at short distances through
tumbling.
Also shown in \fref{stat_corr_func_partial}(a), the distribution $\Pmp$
has a single such maximum within the periodic domain that defines the accumulation distance $\xA$,
the characteristic distance at which anti-aligned particles accumulate.
Comparing $\xA$ with the jamming distance, \fref{jammingDist}(b),
we find that if $\xJ$ exists, it is a lower bound of $\xA$
because
reorientations 
effectively shift the accumulation distance well away from $\xJ$.
Moreover, the crossover from effective repulsion (dots)
to attraction (solid lines) has no obvious relation to the existence of $\xJ$.

\emph{Effective potential} --- Although the system is out of equilibrium,
it can be insightful to express the pair correlations $\Pmp$ through an {effective potential}
$\Weff = -\ln(2L^2\Pmp)$ 
 \cite{HansenMcDonald:2006,
 SlowmanETAL:2016,
 FodorETAL:2016}.
We identify three regions in the effective potential
$\Weff$, shown in \fref{stat_corr_func_compare}(c):
soft repulsion at short distances, 
a deep attractive well at intermediate positive distances,
and no interaction at long distances.
The range and finiteness of $\Weff$ in the repulsive component is unique
to soft interactions  \cite{BroekerETAL:2024,SlowmanETAL:2016}.
We find that, as the activity $\Pe$ is increased, both the repulsive
and the attractive components of $\Weff$ 
are squeezed into a shorter range,
consistent with the findings of Ref.~\cite{FarageETAL:2015}.

In conclusion, our microscopic theory shows how effective attraction
emerges as a higher order effect from soft repulsion between self-propelled particles.
This resolves the link between particle dynamics 
and accumulation, a key ingredient in phase separation.
Our work highlights the power of particle field theories in active matter
and nonequilibrium physics,
akin to Feynman's QED \cite{Feynman:1949}, for its systematicness and accuracy.
This work is a stepping stone towards a microscopic description of 
MIPS
in higher dimensions and larger particle numbers.

RG-M was supported in part by the European Research Council under the EU's Horizon 2020 
Programme (Grant number 740269), 
and acknowledges support from a St John's College Research Fellowship, University of Cambridge.

\bibliography{articles,books}

\end{document}